\documentclass[twocolumn,aps,10pt,prb,superscriptaddress]{revtex4-2}
\usepackage[usenames,dvipsnames]{color}

\usepackage{comment}
\usepackage{graphicx}
\usepackage{inputenc}
\usepackage{lineno}
\usepackage{color}
\usepackage{hyperref}
\usepackage{amsmath}
\usepackage{braket}
\usepackage{caption}
\usepackage{bm}
\usepackage{subcaption}
\usepackage{tikz}
\usetikzlibrary{calc}
\usepackage{mathrsfs}

\usepackage[]{todonotes}

\begin{document}

\title{Feynman diagrammatics based on discrete pole representations: A path to renormalized perturbation theories}
\author{Daria Gazizova}
\affiliation{Department of Physics and Physical Oceanography, Memorial University of Newfoundland, St. John's, Newfoundland \& Labrador, Canada A1B 3X7} 
\author{Lei Zhang}
\affiliation{Department of Physics, University of Michigan,
Ann Arbor, Michigan 48109, United States of America}
\author{Emanuel Gull}
\affiliation{Department of Physics, University of Michigan,
Ann Arbor, Michigan 48109, United States of America}
\author{J. P. F. LeBlanc}
\email{jleblanc@mun.ca}
\affiliation{Department of Physics and Physical Oceanography, Memorial University of Newfoundland, St. John's, Newfoundland \& Labrador, Canada A1B 3X7}

\date{\today}
\begin{abstract}
 By merging algorithmic Matsubara integration with discrete pole representations 
 we present a procedure to  generate fully analytic closed form results for impurity problems at fixed perturbation order.
   To demonstrate the utility of this approach we study the Bethe lattice and evaluate the second order self-energy for which reliable benchmarks exist.  We show that, when evaluating diagrams on the Matsubara axis, the analytic sums of pole representations are extremely precise.  We point out the absence of a numerical sign problem in the evaluation, and explore the application of the same procedure for real-frequency evaluation of diagrams.  We find that real-frequency results are subject to noise that is controlled at low temperatures and can be mitigated at additional computational expense.  
   We further demonstrate the utility of this approach by evaluating dynamical mean-field and bold diagrammatic self-consistency schemes at both second and fourth order and compare to benchmarks where available.
\end{abstract}

\maketitle

\section{Introduction}
The Matsubara formalism for Feynman diagrammatics is a cornerstone of computational condensed matter physics that appears throughout the literature on finite temperature methods for computing many-body systems.  The vast majority of applications use only at the lowest order diagrams such as those for polarization, random-phase approximation, $GW$, and the second-order self-energy. These key low order diagrams find application in a wide range of quantum chemistry and materials physics problems.  Not surprisingly there are many problems for which low order diagrams are insufficient and this leads to renormalized perturbative approaches and self-consistent schemes in an attempt to capture essential features of higher order diagrams.\cite{zgid:sc, rusakov:2019, iskakov:2020, zgid:2017, qpscGW, kresse:scGW, kresse:gw, scgw, Backhouse:2021, harsha2024}  

Moving beyond low orders is rarely straightforward.  However, in the case of non-interacting problems recent advancements allow one to automatically evaluate internal temporal integrals using algorithmic Matsubara integration (AMI)\cite{AMI}  which produces expressions analogous to those typically used at low orders and has been applied to compute a variety of properties of the Hubbard model\cite{taheri:2020,mcniven:2021,mcniven:2022,farid:2023,Gazizova(2023),burke:2023} as well as the correlation exchange kernel in the jellium model.\cite{leblanc:2022,igor:spectral, perdew}
  Using AMI  relies on contour integration methods to evaluate nested Matsubara sums and therefore hinges on having access to the analytic pole structure of a diagram's constituents which typically means that the states (or bands) have known analytic form or energies that can be represented in a diagonal basis.\cite{Assi,jia:2020} This prevents the method from being used for iterative or self-consistent calculations in most cases.  One path around this is to use a full spectral decomposition of a diagram for which the internal integration space grows and becomes quickly intractable as order increases, again limiting evaluation to low-order diagrams.\cite{igor:BS} 

In this paper we explore the use of pole decompositions on Matsubara axis Green's functions and test their utility for evaluating convolutions of Green's functions that arise in Feynman diagrams.  Using the semi-circular density of states as the simplest possible test case we invoke discrete pole decompositions of the Matsubara axis Green's function: the discrete Lehmann representation (DLR), and the Prony representation.\cite{kaye:dlr,prony}  We then evaluate the second-order self-energy using both the full spectral representation as well as derive an analogous expression for the self-energy based on either pole decomposition.  
We find that results for the self-energy employing a discrete pole representation are extremely stable and  accurate on the Matsubara axis and can be obtained at a small fraction of the computational expense when compared to the full spectral representation.  We show how the method can be applied iteratively for data on the Matsubara axis and show that the converged result can be directly evaluated near the real-frequency axis with a finite regulator.  We find that doing so suffers from oscillatory errors due to discretization and suggest methods of mitigating this issue. Finally, we show the method can be extended to arbitrary orders, demonstrate a self-consistency loop performed at fourth order, and highlight differences from lower order self-consistency.

\section{Methods}
\subsection{Standard Spectral Representation}
Feynman diagrams are comprised of products of Green's functions with arguments both internal and external to the diagram.  For simplicity, we will restrict discussion to a self-energy diagrammatic expansion at order $m$.  Each m\textsuperscript{th} order  diagram must be summed over $m$ independent internal Matsubara frequencies, 
$\{i\nu_i\}=(i\nu_1, ..., i\nu_m)$, but contains a product of $2m-1$ Green's functions; $m$ of which have independent labels in $\{i\nu_i\}$, and the remaining $m-1$ have dependent labels that are some linear combination of $\{i\nu_i\}$ as well as an external Matsubara frequency, $i\nu_x$.  We annotate such a 
linear combination as $\vec{\alpha}\cdot \vec{\nu}$ where $\vec{\nu}=(i\nu_1,...,i\nu_m, i\nu_x)$ and $\vec{\alpha}$ is a vector of the same length representing coefficients with values $\pm 1,$ or $0$.  

An arbitrary self energy diagram will take the form
\begin{equation}\label{eq:sigma}
    \Sigma(i\nu_x)=\sum\limits_{\{i\nu_i\}}H(\{i\nu_i\},i\nu_x)
\end{equation}
where $H(\{i\nu_i\},i\nu_x)$ is a product of Green's functions given by
\begin{equation}\label{eq:H}
    H(\{i\nu_i\},i\nu_x)=G(i\nu_1)...G(i\nu_m)G(\vec{\alpha}_1 \cdot \vec{\nu})...G(\vec{\alpha}_{m-1}.
\cdot \vec{\nu}) 
\end{equation}
After a Matsubara sum is performed each term will also be in the form of Eq.~(\ref{eq:H}) allowing the Matsubara sums to be performed in sequence. 

The spectral representation for a single Green's function is an integral representation given by
\begin{equation}\label{eq:standard}
    G(i\nu_n)=\int\limits_{-\infty}^{\infty}\frac{A(x)}{i\nu_n - x} dx,
\end{equation}
where the spectral density, $A(x)$, is a function of a real-frequency, $x$, and in the case of causal functions is related to the imaginary part of the Green's function on the real-frequency axis. 
One can then apply the spectral representation of Eq.~(\ref{eq:standard}) to each Green's function in Eq.~(\ref{eq:H}) and this is the typical spectral representation of a Feynman diagram which, dropping overall prefactors and interaction terms, would appear in the form
\begin{align}\label{eq:startspectral}
    \Sigma(i\nu_x)=&\sum\limits_{\{i\nu_i\}}\iint A(x_1)... A(x_{2m-1}) \\
    & \times \prod\limits_{\ell=1}^{m} \frac{1}{i\nu_\ell-x_\ell} \prod\limits_{j=1}^{m-1} \frac{1}{\vec{\alpha_j}\cdot \vec{\nu}-x_{j+m}}  dx_1 ... dx_{2m-1}.
\end{align}
This is useful because the order of Matsubara integration and spectral integration can be interchanged, and the Matsubara sums can be performed analytically since the product of kernels is identical to a product of non-interacting Green's functions.  The Matsubara sums for any such diagram can be symbolically generated via algorithmic Matsubara integration (AMI) for which computational tools exist.\cite{AMI,libami,torchami}  The result of those summations will be in the form,
\begin{align}\label{eq:sigmaspec}
    \Sigma(i\nu_x)=\iint&  A(x_1)... A(x_{2m-1}) \\ 
    &\times I(\beta,\{ x_i\},i\nu_x)  dx_1 ... dx_{2m-1} \nonumber 
\end{align}
where $I(\beta,\{ x_i\},i\nu_x)$  is given by
\begin{equation}\label{eq:form}
    I(\beta,\{ x_i\},i\nu_x)=\sum\limits_{\text{terms}} F(\beta,\{ x_i\})\prod G_0(i\nu_x,\{ x_i\}). 
\end{equation}
Here $F(\beta,\{ x_i\})$ is an analytic function containing sums and products of Fermi and Bose distribution functions and their derivatives, and $G_0$ are non-interacting Green's functions arising from the kernel of the spectral representation, and the sum is over terms that typically factorize at intermediate steps in the Matsubara summation.  The form of $I(\beta,\{ x_i\},i\nu_x)$ will be distinct for each diagram topology but can be automatically generated by the AMI procedure.\cite{libami,torchami}

This general spectral representation can be used for any choice of spectral density, and so is not restricted to non-interacting problems while the Matsubara sums are still trivially constructed using AMI or, in the case of Green's function expansions, via other direct analytic approaches.\cite{jaksa}  Most importantly, the analytic continuation of the external frequency $i\nu_x\to \omega+i0^+$ is allowed after Matsubara sums are performed.  
However, the limitations of this approach are severe.   One has added to the original problem of summing $m$ independent Matsubara frequencies a new set of $(2m-1)$ nested integrals with infinite bounds.  In many cases the product of spectral functions is very sparse, leading to a  numerical sign (phase) problem when using stochastic methods that typically worsens as order increases.  For that reason, this is typically only done for lowest-order diagrams such as a particle-hole bubble, $\Pi=GG$, or the GW approximation, $\Sigma=GW$, since the number of spectral integrals in these cases is only two, and these can be further reduced to just two 1-dimensional integrals.  
Finally, the form of Eq.~(\ref{eq:form}) holds and can be analytically continued for 2-point correlation functions where there is a single external frequency.
 In the case of multi-point correlators there are multiple Matsubara frequencies that need to be analytically continued and this complication is non-trivial to resolve.\cite{anxiang:2024,anxiang:2024:2}
\subsection{Discrete Pole Representations}
\subsubsection{Discrete Lehmann Representation}
The discrete Lehmann representation (DLR)\cite{kaye:dlr,kaye:libdlr,kaye:3point} is a pole representation\cite{wallerberger:overcomplete,sparse-ir,shinaoka:2022,hiroshi:sparse,shinaoka:jpsj} for functions that have a spectral decomposition that can be well approximated in a truncated form
\begin{equation}\label{eq:spec}
    G(i\omega_n)=\int\limits_{-\Lambda}^{\Lambda} K(i\omega_n,x)\rho(x) dx,
\end{equation}
where $\rho$ is the density and the kernel is given by $K(i\omega_n,x)=\frac{1}{i\omega_n-x}$ when represented in Matsubara frequencies.
The formulation of DLR in the frequency domain assigns a set of frequencies along the imaginary axis, $\{i\omega_{DLR}\}$, that are determined uniquely for a given choice of $\Lambda$ and an error tolerance at a temperature $\beta$ to ensure that the kernel is well approximated in the domain $[-\Lambda,\Lambda]$ based on a discrete set of poles along the real axis, $\{\omega_k \}$.
 If this can be accomplished then the Green's function can be approximated as a sum over a finite set of poles $\{\omega_k\}$ with weights $g_k$ as
\begin{equation}\label{eq:gdlr}
    G(i\omega_n) \approx G_{DLR}(i\omega_n)=\sum\limits_{k=1}^{r} K(i\omega_n, \omega_k) g_k.
\end{equation}
In essence the intent is to leverage the flexibility in defining $\rho(x)$ in Eq.~(\ref{eq:spec}) and is equivalent to assuming 
\begin{equation}\label{eq:rho}
    \rho(x)=\sum\limits_{k=1}^{r} g_k \delta(x-\omega_k).
\end{equation}
In the limits of both $r,\Lambda\to \infty$, the weights should be continuous and if $g_k\to A(\omega_k)$ then the normal spectral representation is recovered. 

\subsubsection{Prony Approximation}
The Prony approximation\cite{prony} results in a pole decomposition of the form 
\begin{equation}\label{eq:gprony}
    G_{P}(z)=\sum\limits_{k=1}^{r}  \frac{g_k}{z-\omega_k}.
\end{equation}
The fit to the function is generated by first mapping an interval of the non-negative imaginary axis to a unit circle.  The Prony approximation then generates a set of poles and weights to satisfy $G(i\omega_n)$.  There are two variants of the Prony approximation we will discuss.    The first, which we call Prony analytic continuation (PronyAC), is the Prony representation as presented in Ref.~\cite{prony} (see also \cite{ying:prony} and \cite{ying:prony:2}) that can be used on discrete data for $G(i\omega_n)$ as a form of analytic continuation to the real frequency axis.    The PronyAC method generates a set of complex poles $\omega_k$ in the lower-half plane with complex weights $g_k$.  The resulting function is valid for arguments $z$ in the entire upper half plane \emph{and} along the real axis.  In this work we use PronyAC as a form of numerical analytic continuation which for sufficiently high-quality data on the Matsubara axis should provide a causal analytic continuation. 

For the purposes of evaluating Feynman diagrams, the function generated by PronyAC is not straightforward to implement, since the function generated by PronyAC only approximates the retarded Green’s function in the upper half of the complex plane but does not approximate the advanced Green’s function in the lower half of the complex plane.  In this paper, we modify the holomorphic mapping in the Prony formalism to generate a symmetrized function valid in both the UHP and LHP resulting in a function with poles that are real, as well as pairs of complex poles that are complex conjugates.  This introduces poles directly on the real axis, similar to DLR.  Nevertheless, the Prony approximation generates a minimal pole representation of the symmetrized function that can then be employed interchangeably with DLR since they are valid in both upper- and lower-half planes.  For the purposes of performing Matsubara sums, we will use this second form of the Prony approximation, and approximate the Green's function on the Matsubara axis with the, $G(i\omega_n)\approx G_P(i\omega_n)$.  

\subsubsection{Analytic Continuation of Diagrams Involving Pole Representations}
Above we have introduced notation for the DLR and the symmetrized Prony Green's functions, $G_{DLR}$ and $G_{P}$ respectively.  Much of the discussion in the following sections pertains to either representation.  We will use $G_D$ to represent either discrete pole representation.
Given a single-particle Green's function on the Matsubara axis, the density is related to the imaginary part of the analytic continuation of the Green's function, $\rho(\omega)=-\frac{1}{\pi}ImG(i\omega \to \omega+i0^+)$.  It is straightforward to see from the discrete nature of the pole representations that we cannot analytically continue the discretized $G_{D}$ and have any hope of reconstructing the original continuous function on the real frequency axis.  This is not surprising since no information regarding the function on the real frequency axis is required to generate a pole representations.  It is therefore the case that $G(\omega+i0^+) \neq G_{D}(\omega+i0^+)$ for any finite number of poles.

However, when evaluating the analytic continuation in practice one includes a finite regulator $i\omega_n \to \omega+i\Gamma$ which should produce a sufficiently accurate result for a sufficiently small choice of $\Gamma$.  We therefore expect that  $G(\omega+i\Gamma) = G_{D}(\omega+i\Gamma)$ for a sufficiently large $\Gamma$, and given the constraint along the imaginary axis that $G(i\omega_0)=G_{D}(i\omega_0)$ (to a controllable precision) we are guaranteed that the pole representation will produce the correct function at low energies for $\Gamma \approx \pi/\beta$.  Therefore, as temperature decreases towards zero $\Gamma$ can be made arbitrarily small. In the case of a discrete representation of a single Green's function on the Matsubara axis, one would not analytically continue using DLR or the symmetrized Prony Green's functions.  These are by construction not well behaved on the real axis.  Instead other methods are preferable such as Nevanlinna\cite{nevanlinna,nevanlinna:code1,nevanlinna:code2} or the above mentioned PronyAC.\cite{prony}

 In what follows we are interested in using pole representations to evaluate sums over products of $G_{D}$ Green's functions.  The above arguments lead us to expect  that such diagram evaluations can be analytically continued directly so long as a sufficiently large regulator is employed, where large is defined by the scale of $\pi/\beta$.

\subsubsection{Matsubara Sums of Products of $G_D$}
When given a Green's function, or product thereof, on the Matsubara axis with a set of known poles, the Matsubara sums can be performed analytically via contour integration.  For example, to treat the sum over a particular function $H(i\omega_n)$ one solves the auxiliary problem of integrating $H(z)$ multiplied by a Fermi or Bose function to match the statistics of $i\omega_n$.  This is then integrated along two contours, C1:along the real axis closed in the upper half plane and C2: along the real axis in the negative direction and closed in the lower half plane. These contours enclose the poles of the Fermi or Bose functions ($f(z)$ and $n(z)$ respectively) which are the set of Matsubara frequencies $\{i\omega_n\}$ as well as the set of poles, $\{z_0\}$, of $H(z)$.  In the case of Fermions, the sum of C1 and C2 give
\begin{align} \label{eq:cont}
\medmuskip=0.1mu
\thinmuskip=0.1mu
\thickmuskip=0.1mu
\nulldelimiterspace=0.1pt
\scriptspace=0pt
    \oint\limits_{C1+C2}{\mkern-18mu f(z)H(z)dz}=0=&-\frac{ 1}{\beta}\sum\limits_{i\omega_n}H(i\omega_n)\nonumber \\ &+ \sum\limits_{\{z_0\}} \text{Res}[f(z) H(z)]_{z_0}.
\end{align}
We see right away that the value of $H(z)$ as a continuous function on the real axis does not appear in the right hand equality, and only the poles of $H(z)$ contribute.  It is tempting therefore to replace the representation of $H(i\omega_n)$ with a discrete pole representation,
\begin{equation} \label{eq:hsum}
     \sum\limits_{i\omega_n} H(i\omega_n) \approx \sum\limits_{i\omega_n} H_{D}(i\omega_n),
\end{equation}
where for a sufficiently high precision representation of the Green's function we presume that $H_{D}$ is given by an appropriate product of $G_{D}$ Green's functions.  Since Eq.~(\ref{eq:cont}) holds for both $H(z)$ and $H_{D}(z)$ one can approximate the original Matsubara sum
\begin{equation}
    \frac{1}{\beta}\sum\limits_{i\omega_n}H(i\omega_n) \approx \sum\limits_{\{\omega_k\}} \text{Res}[f(z) H_{D}(z)]_{\omega_k}.
\end{equation}
A similar relation holds for Bosonic frequency summation with replacement of $f(z)\to -n(z)$.

This simple relationship allows for the systematic sequential evaluation of a series of nested Matsubara sums
over frequencies $i\nu_i$ where each has poles given by the sets $\{\omega_{k_i}\}$ as needed in Eq.(\ref{eq:sigma}). Each $k_i$ runs a range of values from $1\to r_i$ as in Eqs.~(\ref{eq:gdlr}) and (\ref{eq:gprony}). This gives a result analogous to Eq.~(\ref{eq:sigmaspec}) 
\begin{align}\label{eq:sigmadlr}
    \Sigma(i\nu_x)\approx & \sum\limits_{k_1,k_2,...,k_{2m-1}} g_{k_1}...g_{k_{2m-1}} \\
    &\times F(\beta,\{\omega_{k_i}\})\prod G_0(i\nu_x, \{\omega_{k_i}\}). \nonumber
\end{align}
We see from this expression that there is a one-to-one correspondence between the discrete spectral amplitudes, $g_{k_i}$, and the standard spectral functions, $A(x_i)$, in Eq.~\ref{eq:sigmaspec}.  The auxiliary integrations over the continuous valued set of real frequencies $\{ x_i\}$ are now discrete summations over the poles $\{\omega_{k_i}\}$.
It is important to note that the $G_0$ functions that appear in Eq.~(\ref{eq:sigmadlr}) do not arise from the discretization of the density $\rho(x)$ but are actually the kernel of Eq.~(\ref{eq:spec}) and therefore have a well defined analytic continuation such that $i\nu_x$ in Eq.~(\ref{eq:sigmadlr}) can be safely analytically continued to the real-frequency axis. 

If this representation of Matsubara sums is accurate for a manageable number of poles then the reduction in complexity for renormalized perturbative problems is enormous even at low orders.  There are two key advantages.  First, it replaces the arbitrary sampling of each integration axis with a fixed scaling of $(r)^{2m-1}$. Second, when performing iterative calculations it bypasses the evaluation of a real frequency axis density, $\rho(x)$, a process that requires generating a dense grid of real frequencies in order to resolve the salient features of a possibly unknown function.  Instead one only needs to evaluate the self-energy along the Matsubara axis in order to generate a next approximation of $G(\{i\omega_{DLR}\})$ or at an appropriate set of Matsubara points in the upper half plane for the Prony approximation. 

All of this advantage, however, hinges delicately on the accuracy of Eq.~(\ref{eq:sigmadlr}) which cannot be clearly estimated based on accuracy constraints of a DLR or Prony fit.  

\subsection{Test Case}
In what follows we test our approach on a Bethe lattice, an impurity-like problem with no momentum dependence and a semi-circular density of states for which there is a known analytic solution.  The semi-circular density of states is given by $A(\omega)=\frac{1}{2\pi t^2}\sqrt{4t^2 -\omega^2 }$ and the Matsubara Green's function in the upper-half plane given by $G(i\omega_n)=\frac{i}{2t^2}(\omega_n - \sqrt{\omega_n^2 +4t^2})$.  In this representation the bandwidth is $4t$ and we operate in units of $t=1$.

\section{Results}

\subsection{Matsubara-frequency axis}
We first explore evaluation of the self-energy on the Matsubara axis.  As a starting point, using the semi-circular density of states we compute the second order self-energy using the full spectral representation of Eq.~\ref{eq:sigmaspec}, $\Sigma^{(2)}_{spec}(i\nu_x)$ for which the Matsubara sums can be performed by hand. 
In the case of DLR, we generate for the semi-circular density of states the Green's function at $\{i\omega_{DLR}\}$ from which the poles, $\{\omega_k\}$, and weights, $g_k$, are determined using \texttt{libdlr}\cite{kaye:libdlr}.  In the case of Prony, we generate for the semi-circular density of states the Green's function at a finite set of Matsubara points in the upper half plane $\{i\omega_n\}$ from which the poles , $\{\omega_k\}$, and weights, $g_k$, are determined using \texttt{PronyAC}, part of the \texttt{Green} package.\cite{green:zenodo, prony, iskakov2024greenweakcouplingimplementationfullyselfconsistent, pronyac:zenodo, pronyac:github}

The second order contribution in the full spectral representation is given by 
\begin{align}
    \Sigma^{(2)}_{spec}(i\nu_x) &= \iiint A(x_1) A(x_2) A(x_3) \nonumber \\ \times &\frac{[f(x_1) - f(x_2)][n(x_2 - x_1) + f(-x_3)]}{i\nu_x + x_1 - x_2 - x_3} 
     dx_1 dx_2 dx_3,
\end{align}
where $f(x)$ and $n(x)$ are Fermi and Bose distribution functions respectively.
The analogous expression making use of the $G_D$ Green's functions is
\begin{align}
    \Sigma_{D}&(i\nu_x) = \sum_{k_1, k_2, k_3} g_{k_1} g_{k_2} g_{k_3} \\ \times & \frac{[f(\omega_{k_1}) - f(\omega_{k_2})][n(\omega_{k_2} - \omega_{k_1}) + f(-\omega_{k_3})]}{i\nu_x + \omega_{k_1} - \omega_{k_2} - \omega_{k_3}}. \nonumber    
\end{align}
Both  $\Sigma_{\text{spec}}$ and $\Sigma_{D}$ are presented here assuming only simple poles.  In both cases when poles coincide the expressions produce removable divergences and can be treated properly using AMI. For brevity we forego these details.

\begin{figure*}
\center{\includegraphics[width=1.0\linewidth]{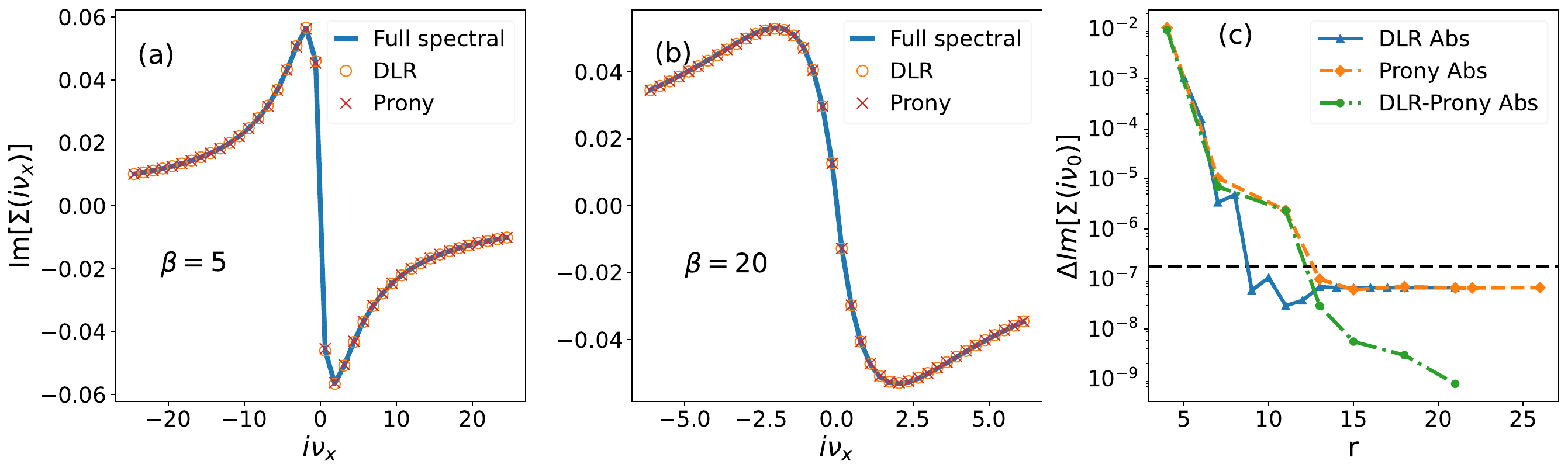}}\\
\caption{Imaginary part of self energy in Matsubara frequencies $i\nu_x$ for the full spectral, DLR and Prony representations. 
(a) $\beta = 5$, (b) $\beta = 20$, (c) Absolute and relative difference for $i\nu_0$ at $\beta=5$.  Relative difference is given by $\Delta\Sigma/\Sigma_{\text{spec}}$. Horizontal dashed line represents the uncertainty estimate of $\Sigma_{\text{spec}}$. 
\label{fig:dlrsigmamat}}
\end{figure*}

For comparison, we evaluate $\Sigma_{D}$ and $\Sigma_{\text{spec}}$ and plot the imaginary parts in Fig.~\ref{fig:dlrsigmamat} (a) and (b) for $\beta=5$ and $\beta=20$ respectively. The analytic sums $\Sigma_{DLR}$ and $\Sigma_{P}$ are visually identical to the integral of the spectral representation for all Matsubara points.  This result is extremely stable to variation in the discrete pole representation, and we scrutinize the accuracy of the comparison in Fig.~\ref{fig:dlrsigmamat}(c).  We plot the absolute deviation of DLR and Prony pole representations from the benchmark spectral result, $\Delta \Sigma = |\Sigma_{D} -\Sigma_{\text{spec}} |$, as a function of the number of poles, $r$, for the zeroth Matsubara frequency.  We also show the deviation $\Delta \Sigma = |\Sigma_{DLR} -\Sigma_{P} |$ to illustrate the behaviour of the pole representations relative to eachother. 
We see that with respect to the $\Sigma_{\text{spec}}$ benchmark both the DLR and Prony representations converge to the same result, with a deviation within the error estimate of the benchmark, $1\times10^{-7}$.  This suggests that both methods are converging to a correct physical value.  One sees that the deviation between the two pole methods does not plateau and that continued increase in the accuracy of the pole representation gives agreement with the two approaches on the scale of $1\times 10^{-9}$ by $r=21$. We believe this represents an essentially exact numerical result, that is also the physically correct value. 
 It is worth noting the difference in computational expense between Eq.~(\ref{eq:sigmaspec}) for $\Sigma^{(2)}_{spec}$ and Eq.~(\ref{eq:sigmadlr}) for the pole representations.  Both have nearly identical analytic form and therefore similar computational expense per evaluation of the integrand/summand. Without prior knowledge of $A(x_i)$ the computational expense of the three nested integrals in $\Sigma^{(2)}_{spec}$ is extremely large.
   Our estimate of $\Sigma^{(2)}_{spec}(i\nu_0)$ in Fig.~\ref{fig:dlrsigmamat}(c) was evaluated to an absolute accuracy of $2\times 10^{-7}$ and required on the order of $10^7$ function evaluations. 
   In contrast the pole representations require precisely $r^3$ evaluations. 
   We see that even for a small number of poles the DLR and Prony results are extremely accurate. At $r=5$ the DLR result is accurate within $2\%$ of the spectral result while only requiring $5^3=125$ function evaluations.  This improves exponentially such that by $r=8$ the relative difference is $1\times 10^{-6}$ for $8^3=512$ evaluations. 

Even for only a second order self-energy, this computational advantage is enormous and will grow exponentially as the diagram order increases.  The pole representations have shifted the spectral integration weight to a finite and \emph{known} set of $\{\omega_k\}$ poles.  This is in contrast to high order integration of continuous functions where an integration routine needs to search for where the integrand has value.  Doing so one encounters sampling issues such as a numerical sign problem.  This is not the case here, where one simply evaluates a discrete set of predetermined points where all the weight has been placed.  There is no-longer a sign problem because there is no stochastic component or variance.  While there remains an exponential expense in the number of poles we foresee opportunities for approximations of the pole representations where one removes poles with small $g_k$ values, but note that such exclusions should be performed with care since in principle their impact would be uncontrolled.

\subsection{Real-Frequency axis}
\begin{figure}
\center{\includegraphics[width=1.0\linewidth]{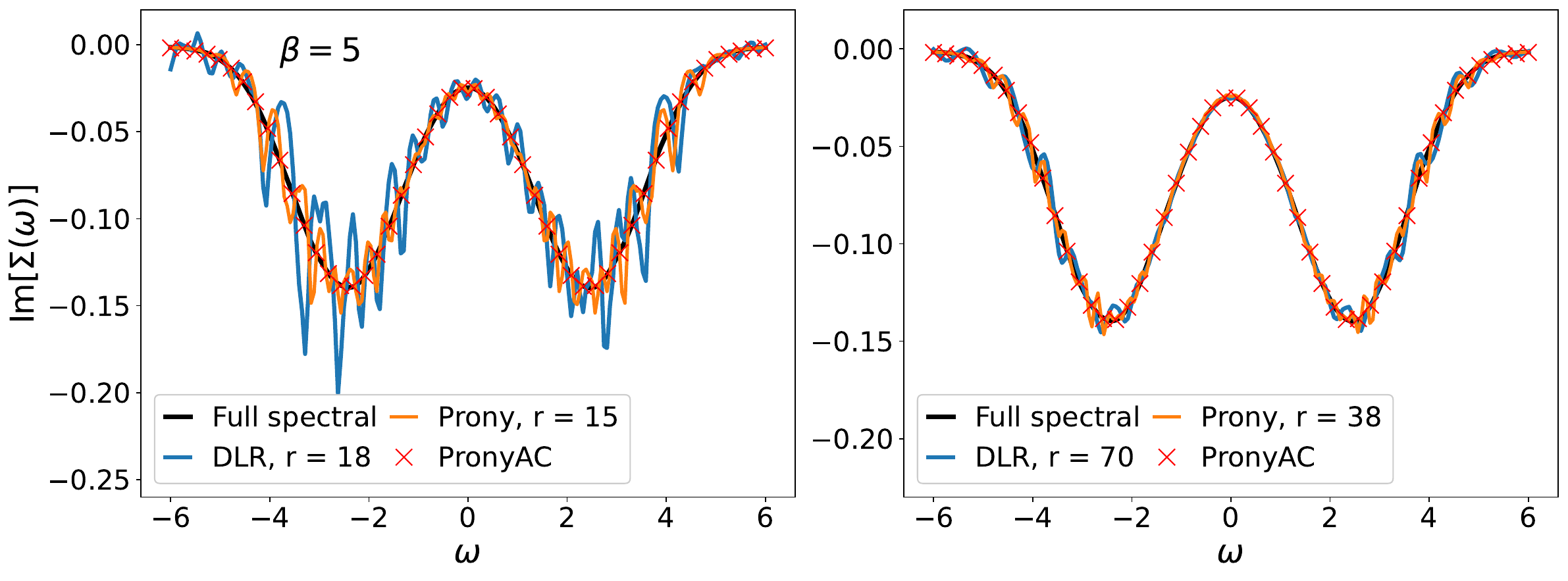}}\\
\vspace{-0.5 cm}
\center{\includegraphics[width=1.0\linewidth]{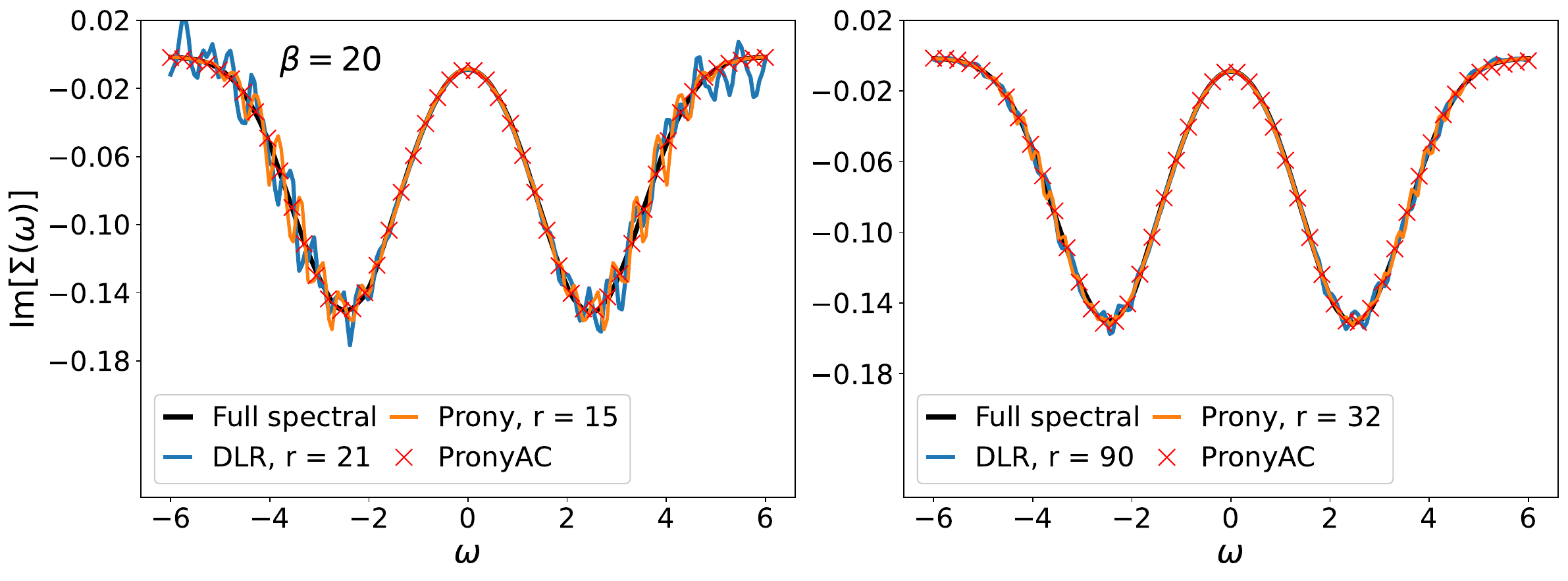}}\\
\caption{Imaginary part of self energy for frequency $i\omega_n\to\omega+i\Gamma$ for $\beta=5$ and $\beta=20$ using spectral, DLR, and Prony representations. Analytic continuation performed using $\Gamma = 0.1$.\label{fig:sigmadlrreal}
}
\end{figure}

Turning to real frequencies, we evaluate ${\rm Im}\Sigma_{\text{spec}}^{(2)}$ and ${\rm Im}\Sigma_{D}^{(2)}$ replacing the external frequency $i\nu_x \to \omega +i\Gamma$ and present results for a choice of $\Gamma=0.1$ for $\beta=5$ and $20$ in the first and second rows of Fig.~\ref{fig:sigmadlrreal} respectively.  One notes that the Prony and DLR results oscillate around the benchmark from the full spectral case.  These oscillations have a number of origins, including overfitting, but are largely due to the fact that $G_{D}$ is not the physical function on the real axis, but has been discretized, and the oscillations originate from the spacing of the real-frequency poles.   

We see from the left hand column of Fig.~\ref{fig:sigmadlrreal} that for a fixed choice of the regulator $\Gamma$ the region at small $\omega$ 
for $\beta=5$ has severe oscillations and that these are dampened by decreasing temperature to $\beta=20$.  
In the case of DLR, we 
see that the oscillations that are absent at low frequency return at higher frequency.  
In the case of the Prony approximation the result is well behaved both at low and high frequencies and exhibits oscillatory behaviour at intermediate points. 
Importantly, in all cases we find that these oscillations are not fundamental to the evaluation, meaning that distinct pole representations will find distinct oscillatory patterns, and therefore these oscillations can be mitigated in a number of ways.  For example, since all pole representations are valid we can generate quantities as averages over distinct DLR/Prony representations to reduce the oscillations.  Another obvious way to reduce the noise is to increase the number of poles in the pole representation by merging multiple DLR/Prony representations to generate finer frequency grids.  This is guaranteed to work since an average over a set of DLR/Prony representations is also a valid pole representation.  

We do precisely this in the right-hand column of Fig.~\ref{fig:sigmadlrreal}.  At fixed $\Gamma$ and $\beta$ we merge a number of DLR/Prony representations and in each case see strong reduction in oscillations on the real-frequency axis.  Finding an optimal way to accomplish this would be an important contribution in the future.  To summarize, results in real frequency are more sensitive to the details of the pole representation, but appear controllable by increasing the number of DLR/Prony nodes. 

What is truly desired is analytic continuation in the $\Gamma\to0^+$ limit, which the above approach cannot accomplish, though it remains a useful approach to check other forms of numerical analytic continuation.  We illustrate also in Fig.\ref{fig:sigmadlrreal} the analytic continuation of PronyAC with the same regulator, which very accurately reproduces the benchmark spectral result with no spurious oscillations.  The causal nature of PronyAC makes it a powerful numerical analytic continuation tool when combined with the high precision Matsubara axis data generated using our AMI toolset.

\subsection{Self-Consistent Calculations}
There are two obvious paths forward using the pole representations for self-consistent calculations.  The first is  self-consistent perturbation theory, or bold diagrammatics, which we can evaluate to any order but here restrict to second order.  This is therefore equivalent to the well known GF2 method.  The second is dynamical mean field theory using the second-order diagram as a solver.  These calculations can be done using either DLR or the Prony representations.  Since both are controlled and produce virtually identical results on the Matsubara axis we present in this section results using the DLR where we leverage the sparseness of the DLR representation that only requires measurement of $r$ points along the imaginary axis $\{i\omega_{DLR}\}$.

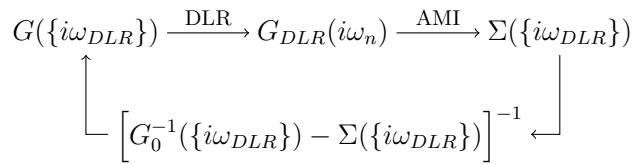
\begin{figure}
    \centering
    \begin{tikzpicture}[node distance=2cm,scale=0.9, every node/.style={scale=0.9}]
\node[fill=white, font=\large] (node1) at (0,0) {$G(\{i\omega_{DLR}\})$};
\node[fill=white, font=\large] (node2) at (3.5,0) {$G_{DLR}(i\omega_n)$};
\node[fill=white, font=\large] (node3) at (7.0,0) {$\Sigma(\{i\omega_{DLR}\})$};
\node[fill=white, font=\large] (node4) at (3.45,-1.5) {$\Big[G_0^{-1}(\{i\omega_{DLR}\}) - \Sigma(\{i\omega_{DLR}\})\Big ]^{-1}$};
\draw[->] (node1) -- (node2) node[midway, above] {DLR};
\draw[->] (node2) -- (node3) node[midway, above] {AMI};
\draw[->] (node3) |- (node4) node[pos=0.75, right] {};
\draw[->] (node4) -| (node1);
\label{fig:Flow chart1}
\end{tikzpicture}
    \caption{Iterative scheme for perturbative calculation.}
    \label{fig:loop}
\end{figure}

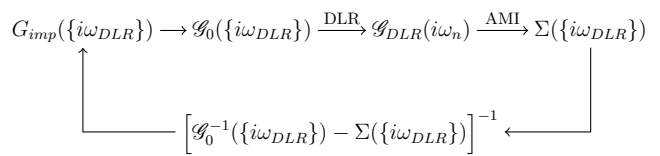
\begin{figure}
    \centering
    \begin{tikzpicture}[node distance=2cm,scale=0.9, every node/.style={scale=0.7}]
\node[fill=white, font=\large] (node1) at (0,0) {$G_{imp}(\{i\omega_{DLR}\})$};
\node[fill=white, font=\large] (node2) at (2.5, 0) {$\mathscr{G}_0(\{i\omega_{DLR}\})$};
\node[fill=white, font=\large] (node3) at (5.0,0) {$\mathscr{G}_{DLR}(i\omega_n)$};
\node[fill=white, font=\large] (node4) at (7.5,0) {$\Sigma(\{i\omega_{DLR}\})$};
\node[fill=white, font=\large] (node5) at (3.8,-1.5) {$\Big[\mathscr{G}_0^{-1}(\{i\omega_{DLR}\}) - \Sigma(\{i\omega_{DLR}\})\Big ]^{-1}$};
\draw[->] (node1) -- (node2);
\draw[->] (node2) -- (node3) node[midway, above]{DLR};
\draw[->] (node3) -- (node4) node[midway, above] {AMI};
\draw[->] (node4) |- (node5);
\draw[->] (node5) -| (node1);
\label{fig:Flow chart2}
\end{tikzpicture}
    \caption{DMFT self-consistency loop.}
    \label{fig:DMFT_loop}
\end{figure}

\subsubsection{Self-Consistent GF2}
\begin{figure}
\center{\includegraphics[width=1.0\linewidth]{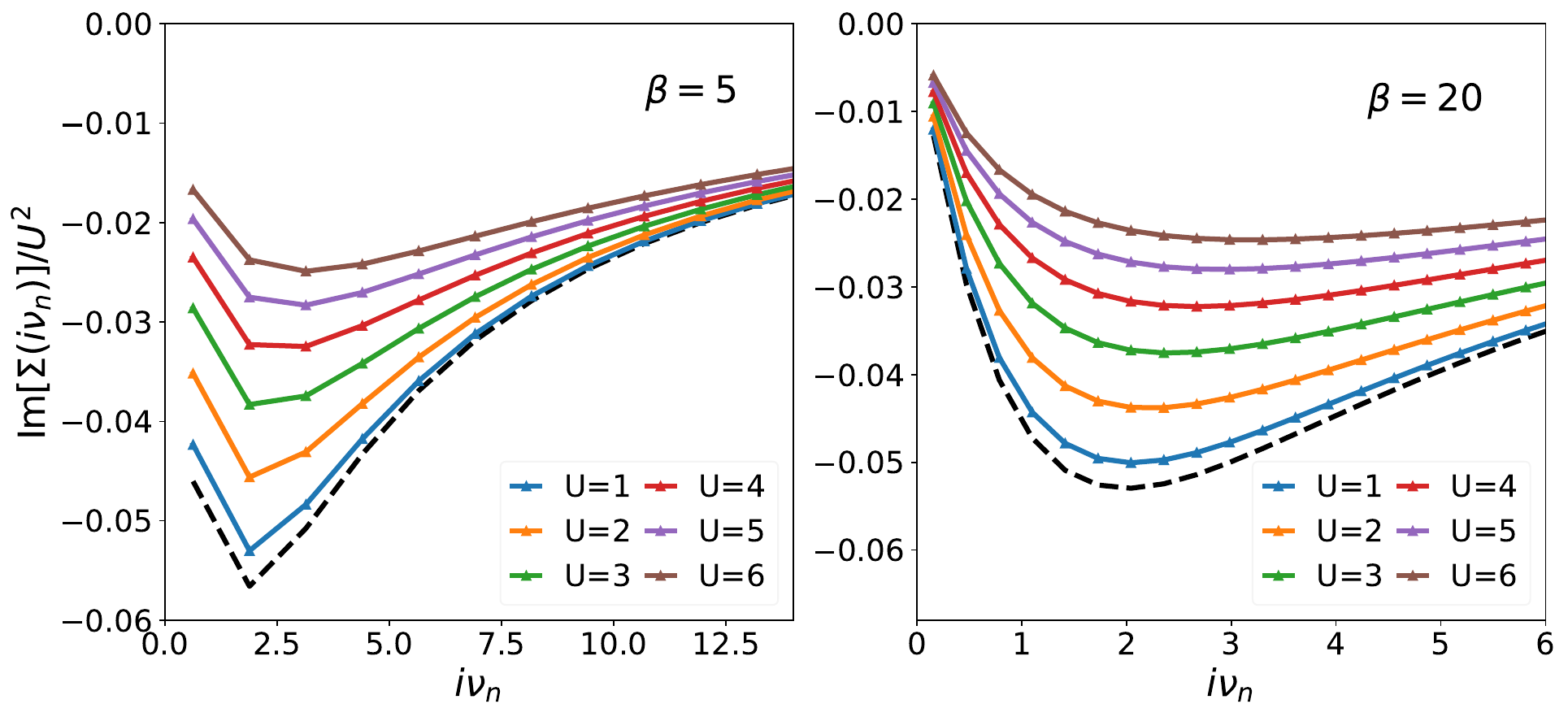}}
\caption{Imaginary part of the GF2 self energy in Matsubara frequencies $i\nu_n$ after self 
consistency procedure for different U values. Left: $\beta = 5$, right: $\beta = 20$. DLR 
representation used $r\approx25$ for $\beta=5$ and $r\approx 40$ for $\beta=20$.
\label{fig:IPTdlrsigmamat} }
\end{figure}

Given a Green's function at the DLR Matsubara points $\{i\omega_{DLR}\}$ we find the DLR representation of the Green's function that is a good estimate at all Matsubara points.  We can then evaluate a self-energy directly at the points needed $\{i\omega_{DLR}\}$ and using the Dyson equation get a new estimate which can then be used to iteratively compute the next estimate of the self-energy.  This iterative scheme is depicted in Fig.~\ref{fig:loop}.
We perform  the self-consistent perturbation theory for second order using the DLR representation and plot the converged results in Fig.~\ref{fig:IPTdlrsigmamat} for $\beta=5$ and $20$ assuming a Hubbard interaction in the range of $U/t=1\to6$  which we suppress in the final result to plot on the same scale.  The overall behaviour of iterative pertubative methods at low orders typically biases the system towards having a metallic characteristic, indicated by the relative difference of the first two Matsubara frequencies\cite{park:2008,fedor:2020}.  Results appear behaved and insensitive to the detailed choice of the DLR representation.  

\subsubsection{DMFT Self-consistency}
Here we explore the use of the DLR representation as a diagrammatic solver for DMFT where benchmark results are readily available.  The self-consistency loop is that of Fig.~\ref{fig:loop} except that in each iteration the Green's function must be replaced with a Weiss field in the form $\mathscr{G}(i\omega_{DLR})=i\omega_{DLR}+\mu-t^2G_{imp}(i\omega_{DLR})$.\cite{georges:revmod,bethe1}  We then find the DLR representation of $\mathscr{G}$ and use this for the next step of the diagrammatic expansion as depicted in Fig.~\ref{fig:DMFT_loop}. 

\begin{figure}
\center{\includegraphics[width=1.0\linewidth]{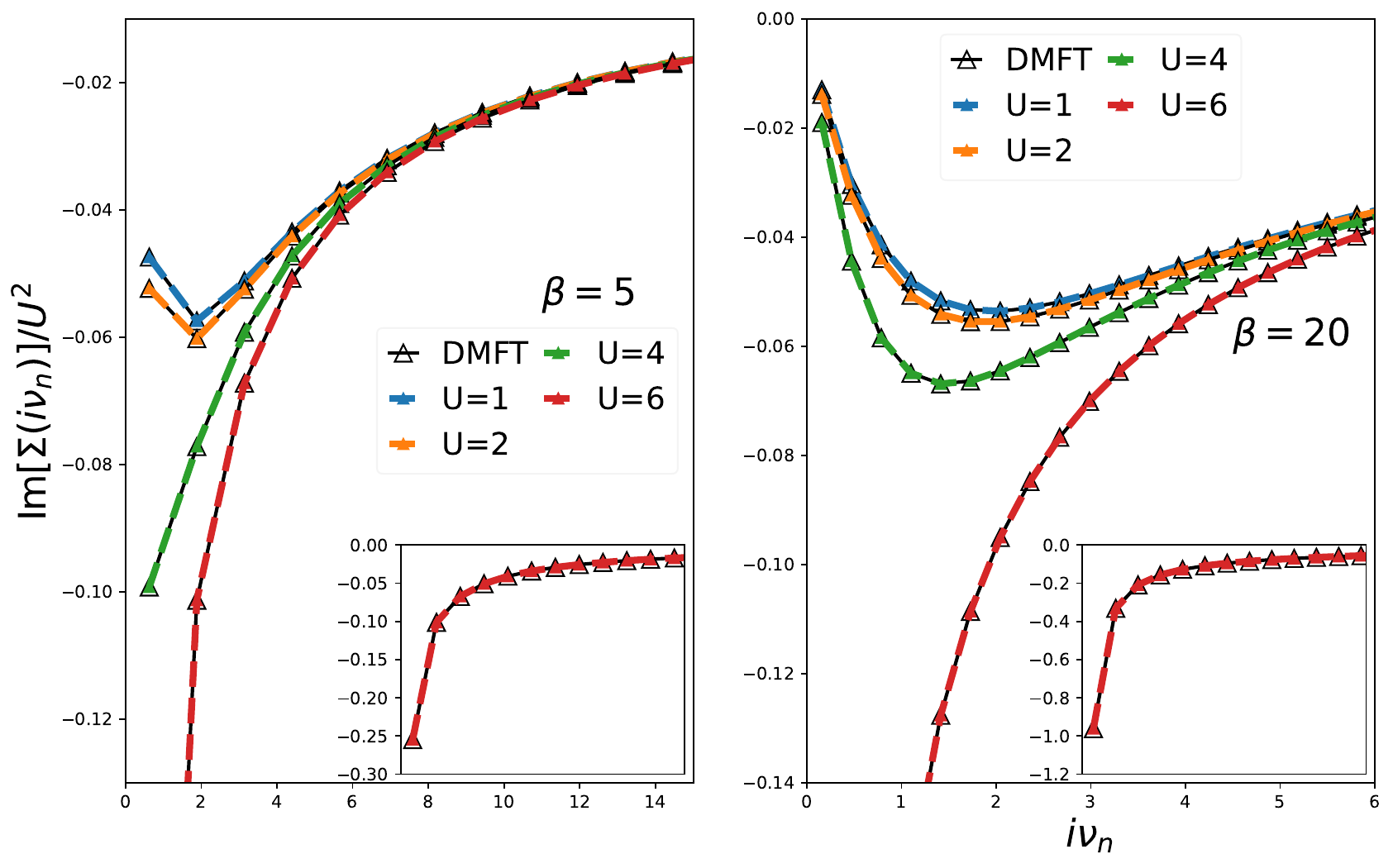}}\\
\caption{Imaginary part of the converged DMFT self energy in Matsubara frequencies $i\nu_n$ for different U values. Left: $\beta = 5$ with $r\approx27$ poles, right: $\beta = 20$, $r\approx40$. Comparison DMFT results generated with TRIQS package.\cite{triqs}
\label{fig:DMFTdlrsigmamat}}
\end{figure}

We compute this first using the DLR representation which requires only the set of $\{i\omega_{DLR}\}$ and then as a benchmark using the standard second-order theory formulated on the imaginary-time axis via the TRIQS package.\cite{triqs}
Results for both are shown in Fig.~\ref{fig:DMFTdlrsigmamat} for $\beta=5,$ and $20$, the DLR using solid-colored points and the  DMFT benchmark using open points.  Since both are performed to second order they are mathematically equivalent and we find perfect agreement between the methods.  There is again a computational advantage at play, the DLR representation requires substantially fewer evaluations for each iteration replacing continuous integrations with a handful of discrete sums, and in this case does not rely on a Fourier transform from imaginary times.  Further, since every DMFT iteration is the evaluation of a closed-form result, there is no stochastic error and we expect our result to be accurate to high precision. 
We note that one could also use the converged self-energy to do a final iteration of Eq.~\ref{eq:sigmadlr} for which we perform analytic continuation $i\nu_n\to \omega+i\Gamma$ using either the DLR or Prony pole representations.  A preferred alternative is to leverage the extreme precision of the DLR+AMI approach on the Matsubara axis to constrain other forms of numerical analytic continuation, such as PronyAC.  Nevertheless, a direct evaluation remains possible as a means of verifying numerical analytic continuation.

\subsection{Beyond Second Order}
\begin{figure}
\center{\includegraphics[width=1.0\linewidth]{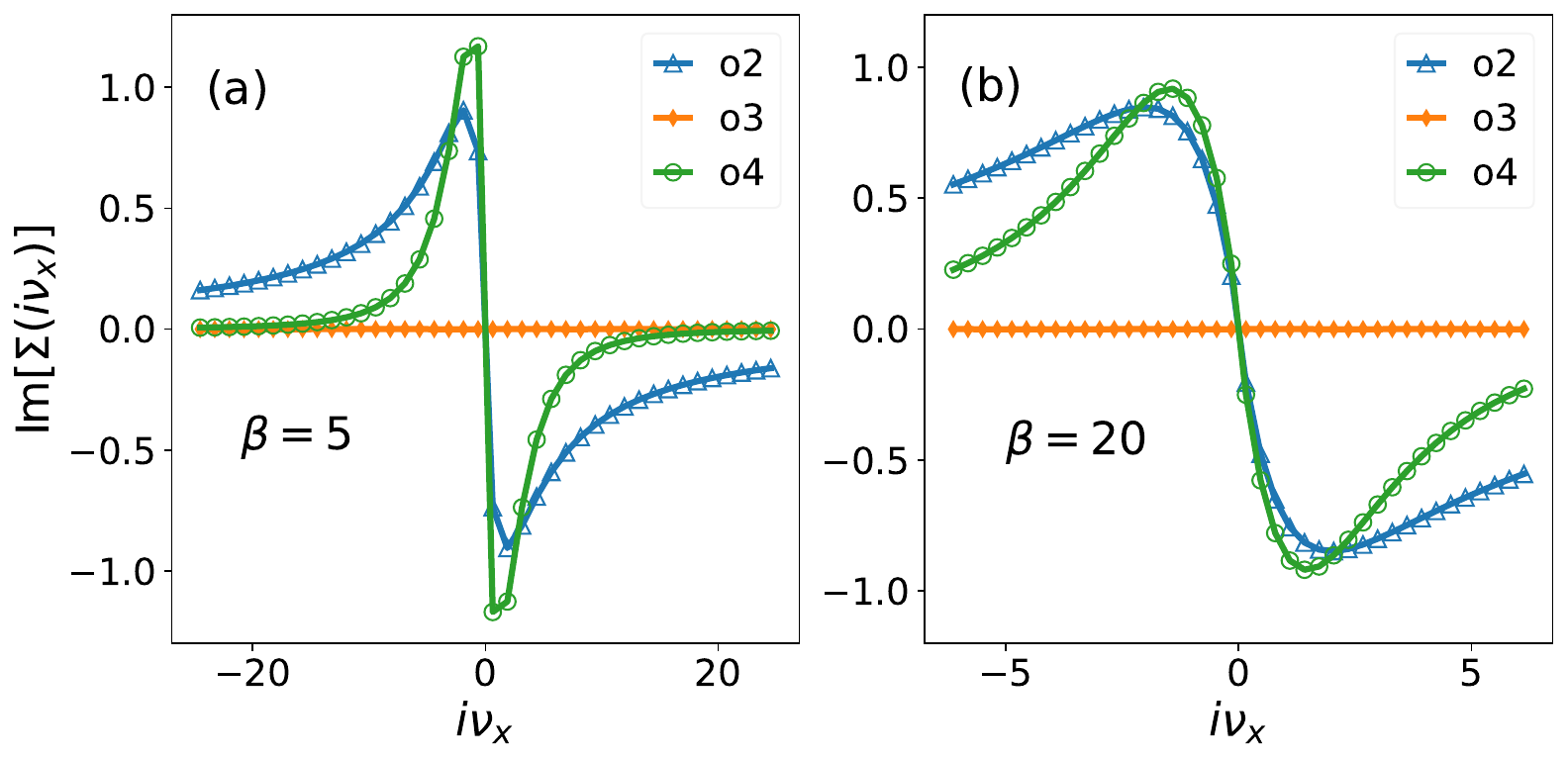}}\\
\caption{Imaginary part of skeleton self energy at orders $2\to4$ in Matsubara frequencies $i\nu_x$ using a DLR representation for $r=14$ at $U = 4$. 
(a) $\beta = 5$, (b) $\beta = 20$. \label{fig:dlrsigmaO4mat}}
\end{figure}

Up to this point we have rigorously shown the utility of pole representations for producing analytic closed form results for second order self-energies, where reliable benchmarks exist, while asserting that this can be used for any diagrammatic expansion.  To support this claim, we repeat the self energy calculations up to fourth order.  This involves evaluating two third-order diagrams, the sum of which is zero for particle-hole symmetric problems such as this, as well as 12 fourth order diagrams.  Of the 12 fourth order diagrams, 3 include self-energy insertions, leaving 9 skeleton diagrams that would be evaluated if doing self-consistency as in Fig.~\ref{fig:loop}.  The DMFT self-consistency would require all 12 fourth order diagrams.  We present results at $U=4$ and $\beta=5$ for the imaginary parts of the second, third and fourth order skeleton series after a single iteration in Fig.~\ref{fig:dlrsigmaO4mat}.  These parameters were selected since the fourth-order is comparable in amplitude to the second-order.  An important check, we see that we recover to high precision the zero contribution from the two third order diagrams, the sum of which returns values on the scale of $1\times 10^{-6}$.  Interestingly, the DLR imposed here does not necessitate more poles as order increases and similar to data for second-order in Fig.~\ref{fig:dlrsigmamat} we find that the result for the imaginary part of $\Sigma^{(4)}(i\omega_0)$  converges by $r=14$.  The reason for this is straightforward.  The pole representation assigns an accuracy target to the single-particle Green's function, and increasing the accuracy requires more poles and also finer resolution of the pole amplitudes.  While it is true that many poles with weights on the order of $1\times 10^{-5}$ are required for high accuracy in the Green's function, since fourth order diagrams include convolutions of seven Green's functions, poles with such small amplitude become irrelevant, and only a small number of heavy weight poles are needed to accurately reflect the higher-order diagrams.  

\begin{figure}
\center{\includegraphics[width=1.0\linewidth]{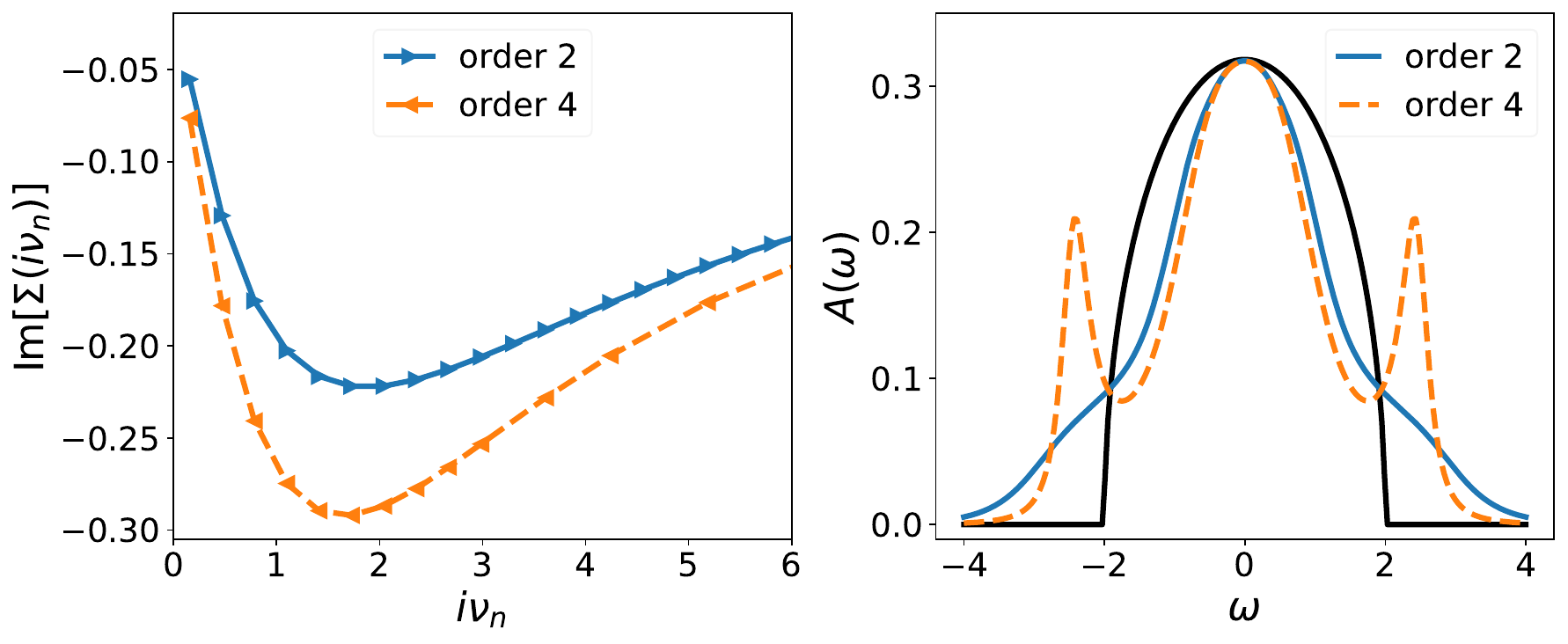}}\\
\caption{Left: Imaginary part of the converged DMFT self energy in Matsubara frequencies $i\nu_n$ generated using DLR representation at $\beta = 20$ an $U = 2$. Right: Spectral function $A(\omega)$ using PronyAC. \label{fig:dlrDMFTo4}}
\end{figure}

With this in mind we perform self-consistency within the DMFT framework including self energies up to fourth order that are iterated to convergence.  We present these in the left-hand frame of Fig.~\ref{fig:dlrDMFTo4} contrasted to the earlier result at second order.  Here we have chosen the weak coupling case of $U=2$ and $\beta=20$.  We see that the corrections at fourth order are not small. While the two curves are qualitatively similar, one should recall that small changes on the Matsubara axis can have substantial impact on real-frequency properties.  To illustrate this, we contrast the real axis results in Fig.~\ref{fig:dlrDMFTo4} right hand frame for the non-interacting, second order DMFT, and fourth order DMFT solutions. These curves represent the analytic continuation with no numerical requlator, $\Gamma=0^+$, enabled by PronyAC.  We see that while low frequency features of $A(\omega)$ are the same, near the band edge, what was a shoulder at second order is now a peak at fourth a hallmark feature of metal-insulator transitions. 

We note that single-shot self-energy calculations at fourth order have been performed in the past.  For example, Ref.~\cite{Gebhard2003} laboriously derives integral expressions for the fourth order diagrams which, for particle-hole symmetric problems, collapses from 12 diagrams to 4 unique sets of 3 diagrams.\cite{werner:2013, GIT}   They extract the imaginary contributions to $\Sigma^{(4)}$ using a partial spectral representation, leaving a dozen or more integrals to be performed.  
In our case, we automatically generate the necessary closed-form analytic expressions using AMI allowing us to evaluate both the real and imaginary parts simultaneously as well as perform self-consistency at minimal computational expense while not being restricted to particle-hole symmetric problems.  
We note that other approaches involving the DLR representation merged with the hybridization expansion show promising results as well\cite{kaye:hybridization}
and have introduced additional simplification that could be adapted to the context of this work.

\section{Conclusions}
We have outlined a path to performing renormalized diagrammatic calculations by generating a pole representation of the Green's function.  We compute GF2 and DMFT self-consistencies for the Bethe lattice using the discrete Lehmann representation combined with algorithmic Matsubara integration. Results are  well behaved and essentially exact on the Matsubara axis.  We also demonstrate that the result can be analytically continued to the real frequency axis  We have combined this approach with algorithmic Matsubara integration which allows us to evaluate any diagram using either a DLR or Prony representation.
To demonstrate this we evaluate the bold diagrammatic resummation up to fourth-order at each iteration (one might call this GF4).  Our results represent an exact and purely analytic result of the self consistency with the only error coming from the expansion into a DLR/Prony basis which is controllable and vanishingly small.
We foresee many opportunities for extending this work. In particular, we note that the procedure is not restricted to impurity problems.  Alternatively, one can take a converged result on the Matsubara axis from a preferred method such as DMFT (including cluster extensions such as DCA\cite{maierrevmod,gullrevmod}, dual Fermions\cite{rubtsov:2009,leblanc:2019}, or dynamical vertex approximation\cite{dGa}), generate a pole representation and then use the $G_D$ Green's function to evaluate \emph{any} diagrammatic expansion of interest on both the real- and imaginary-frequency axes.  This would be particularly impactful for density-density correlation functions and conductivities that cannot be expanded around the non-interacting solution\cite{maxence,Michon2023} and we expect this can also be adapted to problems in quantum chemistry.\cite{Backhouse:2021,zgid:2014:so,backhouse:2020, backhouse:2023,zgid:sc, zgid:thc, jia:2020, Assi}
More generally, merging pole representations of Green's functions, effective interactions, and vertex functions for evaluating Feynman diagrams would open the doors to a wide array of renormalization procedures that might be tuned to specific problems or applications. We note as well that other approaches for real-frequency evaluation exist, and might yield similarly useful results via pole representations.\cite{kugler:realfreq,Ritter2022}
For example, one can merge the DLR representations of bosonic 2-point or 3-point functions as a tool for collapsing the diagrammatic expansions.\cite{kaye:3point,ir,markus,markus2}
We emphasize that Eq.~(\ref{eq:sigmadlr}) is general and can be applied to any problem while similar expressions for polarization and other low order diagrams are trivially converted to the pole representation.  
Finally, we surmise that alternate pole representations might be constructed that 
do not suffer from discretization issues on the real axis\cite{prony} or that 
use as a constraint a smoothness criteria for analytically continued low-order diagrams for a particular choice of $\Gamma$.  Such a representation would allow for optimal evaluation on the real-frequency axis without the irregularities present in this work.

\section{Acknowledgement}
JL acknowledges the support of the Natural Sciences and Engineering Research Council of Canada (NSERC) RGPIN-2022-03882. EG and LZ acknowledge support from NSF QIS 2310182.

\bibliographystyle{apsrev4-2}
\bibliography{refs.bib}

\end{document}